\documentclass[nolineno,a4paper,USenglish,cleveref,nameinlink,autoref]{gd-lipics-v3}

\usepackage{booktabs}
\usepackage[table,dvipsnames]{xcolor}
\usepackage[nocompress,noadjust]{cite}
\usepackage[ruled]{algorithm}
\usepackage{algpseudocode}
\usepackage{xspace}

\theoremstyle{plain}
\newtheorem{problem}{Problem}

\title{How to Draw a Planar Graph: An~Experimental~Evaluation}
\titlerunning{How to Draw a Planar Graph}

\author{Sergey Pupyrev}{Menlo Park, CA, USA}{spupyrev@gmail.com}{https://orcid.org/0000-0003-4089-673X}{}

\authorrunning{S. Pupyrev}

\Copyright{Sergey Pupyrev}

\ccsdesc[500]{Theory of computation~Graph algorithms analysis}
\ccsdesc[300]{Mathematics of computing~Graph theory}
\keywords{Planar graph drawing, drawing aesthetics, experimental evaluation}

\supplementdetails[subcategory={Code and Data}]
  {Software}
  {https://github.com/spupyrev/planar-vibe}

\acknowledgements{}

\hideLIPIcs
\graphicspath{{pics/}}

\definecolor{defblue}{rgb}{0.121,0.47,0.705}
\DeclareTextFontCommand{\df}{\color{defblue}\em}
\definecolor{complexitylow}{rgb}{0.86,0.95,0.86}
\definecolor{complexitymedium}{rgb}{1.0,0.94,0.76}
\definecolor{complexityhigh}{rgb}{0.98,0.82,0.82}
\definecolor{complexityextreme}{rgb}{0.55,0.08,0.08}

\newcommand{\alg}[1]{\textsc{#1}\xspace}
\newcommand{\aes}[1]{\textup{\textsc{#1}}\xspace}

\newcommand{\complow}{\cellcolor{complexitylow}low}
\newcommand{\compmedium}{\cellcolor{complexitymedium}medium}
\newcommand{\comphigh}{\cellcolor{complexityhigh}high}
\newcommand{\compextreme}{\cellcolor{complexityextreme}\textcolor{white}{\textbf{extreme}}}

\EventEditors{Maarten L\"{o}ffler and Silvia Miksch}
\EventNoEds{2}
\EventLongTitle{34th International Symposium on Graph Drawing and Network Visualization (GD 2026)}
\EventShortTitle{GD 2026}
\EventAcronym{GD}
\EventYear{2026}
\EventDate{August 19--21, 2026}
\EventLocation{Ontario, Canada}
\EventLogo{}
\SeriesVolume{396}
\ArticleNo{1}

\begin{document}

\maketitle

\begin{abstract}
    Planar graphs are central to graph drawing, with extensive results on planar
    layouts and related structures. Every planar graph
    admits a planar straight-line drawing, and algorithms can guarantee
    additional geometric or combinatorial properties. However, it is unclear which
    algorithms work best in practice. Even for small graphs with near-perfect manual
    drawings, standard algorithms might produce poor spacing, distorted faces, or
    small angles.

    We present an experimental evaluation of planar graph drawing algorithms on
    a large benchmark collection of small and medium-sized planar graphs
    (\(10\)--\(400\) vertices).
    The study compares established algorithms from the graph drawing literature,
    practical force-directed and pressure-based heuristics, and new
    optimization-based methods that directly improve visual properties such as
    edge-length uniformity, face-area balance, and angular resolution.
    The results show that no evaluated algorithm is best across all aesthetic
    criteria, and optimizing one visual property often worsens another. Directly
    optimizing visual criteria improves targeted scores, and score-guided
    combination of several methods gives the best aggregate results, but no simple
    algorithm emerges as a clear universal default.
    Designing a simple, robust algorithm that performs well across graph families
    and aesthetic criteria therefore remains an open practical problem.
\subparagraph{Generative AI Declaration}
GPT-5.5 was used, under the author's direction, to implement the software used in this study. This software produced the experimental data reported in the paper. The author designed the study, selected the algorithms, benchmarks, and metrics, analyzed and interpreted the results, and takes full responsibility for the implementation and all reported findings.
\end{abstract}

\section{Introduction}

\subsection{Motivation}

Planar graph drawing is one of the central topics in graph drawing, with a long
line of results that enable strong theoretical guarantees.
Classical methods such as Tutte-style embeddings show
that every planar graph admits a planar straight-line drawing without crossings~\cite{Tut63}, and theoretical algorithms guarantee further
geometric properties of the resulting drawing.
This body of work is supported by a rich
combinatorial machinery, including canonical orders, Schnyder woods, and
barycentric representations, making planar drawings one of the
most thoroughly studied problems in the field.

\begin{figure}[!tb]
    \centering
    \includegraphics[width=\linewidth]{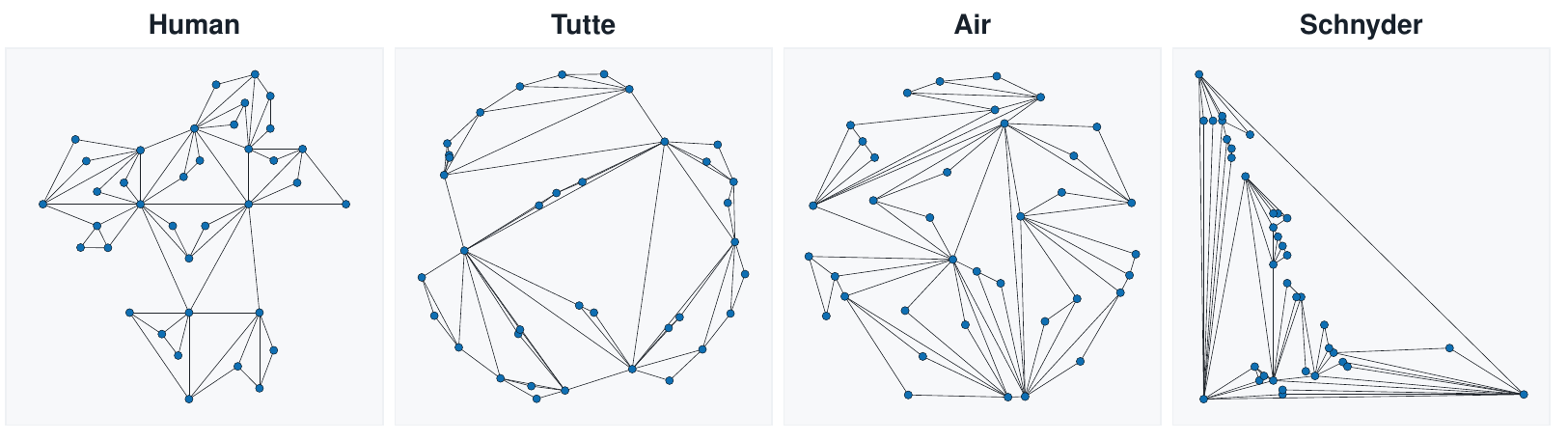}
    \caption{Four planar straight-line drawings of the same graph: a
    human-created layout from the graph-drawing Turing-test
    study~\cite{PAKNPW20}, Tutte's barycentric
    embedding~\cite{Tut63}, the air-pressure method~\cite{Kle18b}, and
    Schnyder's grid-drawing construction~\cite{Sch90}. The human
    drawing is arguably the most visually appealing, and \alg{Air}
    produces a comparably readable layout, while \alg{Tutte} and
    \alg{Schnyder} produce vertex crowding, small angles, and
    non-uniform edge lengths.}
    \label{fig:teaser}
\end{figure}

Despite this depth, it is far less clear which algorithms should be used
to draw planar graphs in practice. The existing literature is largely
theoretical and fragmented along two axes. First, most methods target a
single objective such as convexity of the graph faces~\cite{Tut60,BFM07,CEG23}, areas of the faces~\cite{BRV13,Kle18a,EFKK21}, edge-length ratios~\cite{BF20},
or the size of the underlying grid~\cite{FPP90,Sch90,Kan96,CON85}.
Second, several strong guarantees are often restricted
to specific subfamilies of planar graphs, such as face-area universality for
planar 3-trees~\cite{BRV13}, subcubic graphs~\cite{Tho92} or bipartite graphs~\cite{EFKK21}, and optimal edge-length ratios for series-parallel graphs~\cite{BF20}.
As a result, existing theoretical algorithms often produce layouts with
collapsed faces, small angles, or highly non-uniform edge lengths even on
relatively small instances; see \cref{fig:teaser} for examples.
At the same time, practical heuristics for addressing these defects
are surprisingly difficult to implement and engineer well.
In particular, force-directed and stress-based approaches appear conceptually
simple, but obtaining robust planar straight-line drawings of consistently high
quality turns out to require many design decisions, heuristics, and parameter
choices~\cite{SAAB11,ADDKL22,DLFQIRN09,BRR19}.
Furthermore, empirical evidence suggests that
humans can still distinguish and often prefer hand-drawn layouts to
automatically generated ones even for small graphs~\cite{PAKNPW20}.

The goal of this paper is to make the gap between theory and practice more
visible. We conduct a systematic experimental evaluation of planar straight-line
drawing algorithms.
The evaluation combines classical theoretical algorithms, practical heuristics,
and optimization-based methods developed in this paper under ten aesthetic
criteria capturing spacing, angles, edge lengths, face geometry, and alignment.
Automatic layouts are assessed computationally using these metrics, while human-created layouts provide a comparison; no human-subject
study is conducted.
Rather than proposing a new theory of drawing quality, we aim to identify where
existing algorithms succeed and fail, provide a reproducible open-source
implementation, and formulate concrete open problems motivated by the
evaluation.

\subsection{Contributions}

The contributions of this paper are organized around one question:
which algorithms produce high-quality straight-line drawings of
general planar graphs in practice?

\begin{enumerate}
    \item \textbf{Experimental evaluation.}
    We compare a collection of algorithms for straight-line
    planar graph drawing on a large set of small and medium-sized (\(10\)--\(400\) vertices) planar
    graphs using ten aesthetic metrics. The evaluation studies aggregate
    quality, metric tradeoffs, robustness, runtime, and the differences
    between automatic and human-created drawings. The experiments confirm
    that no single-objective method is uniformly best, and that optimizing
    one aesthetic frequently worsens others.

    \item \textbf{New optimization-based algorithms.}
    We introduce four algorithms for planar straight-line drawing.
    Three, called \df{balancers}, use a common barycentric
    optimization model and tune the drawing toward a chosen aesthetic
    objective: angular resolution, edge-length uniformity, or face-area
    balance. The fourth one, \alg{Hybrid}, is a score-guided method that combines multiple layout strategies and local refinement. It attains
    the highest aggregate metric scores, as expected from its design, which shows the potential of combining methods; however, its high
    implementation complexity and remaining gap to human-created drawings motivate
    future work on simpler general-purpose drawing algorithms and human-aligned
    quality measures.

    \item \textbf{Implementation and benchmark.}
    We present \textsc{PlanarVibe}~\cite{PV26}, an open implementation,
    interactive playground, and benchmark suite for planar straight-line
    drawing, with implementations in \texttt{C++}, \texttt{Python}, and
    \texttt{JavaScript}. \textsc{PlanarVibe} brings together algorithms from
    the literature, practical heuristics, and the new methods developed in
    this paper under a common preprocessing and evaluation pipeline.
\end{enumerate}

\subsection{Related Work}

\subparagraph*{Theoretical Algorithms.}
Planar straight-line drawing has a long history of algorithms with strong
combinatorial and geometric guarantees. Foundational results include
Tutte's barycentric method for convex drawings of $3$-connected planar
graphs~\cite{Tut63,BFM07}, canonical-order and Schnyder-wood constructions
for grid drawings~\cite{FPP90,Sch90,Kan96}, and methods with guarantees on
convexity~\cite{Tut60}, grid size~\cite{CON85}, edge-length
ratios~\cite{BF20}, or prescribed face areas~\cite{BRV13,Kle18a,EFKK21}.
These results are
typically analyzed through one structural property at a time, and
the guarantees often apply only to particular graph classes.
Our focus is different: we evaluate how such methods behave in practice on
general planar graphs under several aesthetic criteria.

\subparagraph*{Practical Methods.}
The most widely used practical layout methods are force-directed,
stress-based, and barycentric approaches. In the planar setting, Tutte-style
barycentric drawings are attractive because positive weights with a fixed
convex outer face preserve planarity and convexity for $3$-connected graphs,
but they often create degenerate drawings that squeeze large parts of the graph
into a tiny region. This problem can be addressed by an iterative adjustment
of edge weights~\cite{FS21} or by a careful assignment of the weights
to improve vertex spread~\cite{CEG23}.
Other practical approaches adapt force-directed
layouts to preserve planarity via movement limits~\cite{SAAB11,Ber00,GBK24}, while air-pressure methods rebalance
faces through local pressure-like updates~\cite{Fel14,Kle18b,KMST25}.
These families are represented in our evaluation and motivate the new
optimization-based methods developed in this paper.

\subparagraph*{Empirical Studies and Aesthetic Metrics.}
Empirical work has compared general graph-drawing
algorithms~\cite{DGLTTV97,ADDKL22}, hierarchical
methods~\cite{BGLPTTVV00}, orthogonal layouts~\cite{KDMW16}, and
user-generated drawings~\cite{DLFQIRN09}; see \cite{DCSPWD24} for a
systematic review. A related line of work
formalizes aesthetic metrics, studies their distributions and correlations,
and considers ways to aggregate them~\cite{HEHL13,MPWK24,MHWMP25,MPK19}.
Many empirical studies identify edge
crossings as one of the most important readability
factors~\cite{Pur97,KPS14}, whereas in our setting
the drawing is required to be crossing-free.
These studies inform the metric framework used in
this paper, but they are not centered on drawings of arbitrary planar graphs.

\subparagraph*{Human Preferences.}
Studies of graph drawing perception show that algorithmic layouts do not
always match human preferences, and that hand-drawn layouts can make
task-relevant structure more explicit~\cite{PPP12,DLFQIRN09}.
We use the human-created drawings from~\cite{PAKNPW20} as part of our benchmark
and compare manual drawings with automatic layouts in \cref{sect:human}.

\section{Experimental Setup}

The experiments presented in this paper were conducted using a new open
source framework, \textsc{PlanarVibe}, that implements existing and new
algorithms in \texttt{C++}, \texttt{Python}, and \texttt{JavaScript}~\cite{PV26}.
Experimental runs were performed on a Linux-based
laptop with a 3.6 GHz Intel Core Ultra 5 125U processor and 16 GB of RAM.
Each algorithm is given the same connected planar input graph and must
produce a planar straight-line drawing.

\subsection{Benchmarks}

We evaluate the algorithms on four collections of connected planar graphs.

\begin{itemize}

\item \textsc{North} is derived from the AT\&T/North DAG collection, which contains graphs extracted from North's Draw DAG service at AT\&T Bell Labs~\cite{BGLPTTVV00}.
From this source, we discard the edge directions and retain the connected planar subset, yielding $854$ graphs.

\item \textsc{Rome} is obtained from Rome-Lib, a standard graph-drawing benchmark~\cite{DGLTTV97}.
Its connected planar subset contributes $3,\!279$ graphs.

\item \textsc{GD-collection} comes from GD-collection-v1, a corpus of 4,890 graph drawings automatically harvested from 27 Graph Drawing conference proceedings (1998--2024), consisting largely of manually created drawings and including many curved or polyline edges~\cite{MHWMP25}.
We keep only those instances that are connected, planar, and already encoded with straight-line edges, resulting in 807 graphs.

\item Finally, \textsc{Named graphs} is a small curated benchmark,
combining the experimental data
of the graph-drawing Turing test study~\cite{PAKNPW20} with
a collection of named Wikipedia graphs. From this source we extract 48 connected planar instances.

\end{itemize}

\noindent
The entire benchmark suite contains $4,\!988$ graphs, with $10$--$400$ vertices
and $9$--$672$ edges; fewer than $1\%$ of the instances are duplicates.

\begin{table}[!tb]
    \centering
    \caption{Summary of the aesthetic metrics used in this paper.}
    \label{tab:metrics-summary}
    \small
    \begin{tabular}{p{0.30\linewidth}p{0.15\linewidth}p{0.45\linewidth}}
        \toprule
        Metric & References & Purpose \\
        \midrule
        \aes{Angular Resolution} & \cite{MHWMP25,ADDKL22,DLFQIRN09,SAAB11} & Rewards large incident angles around each vertex \\

        \aes{Aspect Ratio} &
        \cite{MHWMP25,ADDKL22,DGLTTV97} & Penalizes drawings that are overly stretched in one direction \\

        \aes{Axis Alignment} & \cite{Pur97,PPP12,KDMW16} & Measures reuse of horizontal and vertical coordinate lines \\

        \aes{Convexity} & \cite{Tut60,Tut63,CON85,BFM07} &
        Measures the fraction of bounded faces that are convex \\

        \aes{Edge-Length Deviation} & \cite{MHWMP25,GBK24,ADDKL22} & Rewards edge lengths that stay close to a common target \\

        \aes{Edge-Length Ratio} & \cite{DBETT99,BF20,CEG23} & Measures the gap between the shortest and longest edge \\

        \aes{Edge Orthogonality} & \cite{MHWMP25,PPP12} & Rewards edges close to horizontal or vertical orientation \\

        \aes{Face-Area Uniformity} & \cite{Kle18a,Kle18b,BRV13} & Rewards balanced areas among bounded faces, accounting for face size \\

        \aes{Node Uniformity} & \cite{MHWMP25,MPWK24,DLFQIRN09,MPK19} & Measures how evenly vertices occupy the drawing area \\

        \aes{Spacing Uniformity} & \cite{MHWMP25,CEG23} & Detects local crowding via nearest-neighbor distances \\
        \midrule
        Total Score &  & Arithmetic mean of the ten metric scores \\
        \bottomrule
    \end{tabular}
\end{table}

\subsection{Metrics}

To quantify the quality of drawings produced by various algorithms, we
employ a set of ten \df{aesthetic metrics} that are often discussed in the
context of planar graph drawing. Whenever possible, we reuse the formulas
defined in prior literature, and adapt the definitions when needed. Every metric
is represented by a score in \([0,1]\); larger values are always better.

A summary of the metrics used is given in \cref{tab:metrics-summary}, and the
exact definitions are given in \cref{app:metrics}. The metrics quantify
common layout objectives for planar graph drawings. The spacing metrics
measure whether the drawing uses its bounding box effectively and avoids local
crowding, through \aes{Aspect Ratio}, \aes{Node Uniformity}, and
\aes{Spacing Uniformity}~\cite{MHWMP25,ADDKL22,DGLTTV97,CEG23}. The
orthogonality metrics capture grid-like structure, either by reusing horizontal
and vertical coordinate lines or by making edges close to axis-parallel
directions, through \aes{Axis Alignment} and \aes{Edge Orthogonality}~\cite{PPP12,KDMW16,MHWMP25}.
The angular metric, \aes{Angular Resolution}, rewards large angles between
incident edges~\cite{MHWMP25,ADDKL22,DLFQIRN09,SAAB11}. The
edge-length metrics, \aes{Edge-Length Deviation} and \aes{Edge-Length Ratio},
measure how uniformly edge lengths are distributed~\cite{MHWMP25,GBK24,ADDKL22,CEG23}.
Finally, the face-based metrics, \aes{Convexity} and
\aes{Face-Area Uniformity}, reward convex faces and balanced bounded-face
areas~\cite{Tut60,Tut63,CON85,BFM07,Kle18a,Kle18b,BRV13}. Recall that we
insist on avoiding edge crossings in this study; hence, we do not include the
number of crossings as a metric, although it is arguably the most widely studied
aesthetic criterion in the graph drawing community.

Finally, as a compact diagnostic summary, we report the arithmetic mean of the
ten metric values. This aggregate score is not intended as a model of human
preference.

\subsubsection{Metric Definitions}
\label{app:metrics}

Assume that \(G=(V,E)\) is represented by a planar straight-line drawing, with
\(n=|V|\), \(m=|E|\), and vertex positions
\(p(v)=(x_v,y_v)\in\mathbb{R}^2\). For each edge \(e=\{u,v\}\), let
\(\ell_e=\|p(u)-p(v)\|_2\), and let \(\mathcal{F}\) be the set of bounded
faces.

\subparagraph*{Angular Resolution.}
Incident edges at a vertex should be well separated.
For each vertex \(v\) of degree \(d_v\ge 2\), let
\(\alpha_v^{\min}\) be the smallest cyclic angle between consecutive incident
edges and let \(\alpha_v^\star=2\pi/d_v\) be the ideal angle under equal
spacing. With \(V_2=\{v\in V:d_v\ge 2\}\), define
\[
\df{\aes{Angular Resolution}}
=\frac{1}{|V_2|}\sum_{v\in V_2}
\frac{\alpha_v^{\min}}{\alpha_v^\star}.
\]

\subparagraph*{Aspect Ratio.}
Drawings should avoid being excessively stretched in one direction.
Let \(W=\max_{v\in V}x_v-\min_{v\in V}x_v\) and
\(H=\max_{v\in V}y_v-\min_{v\in V}y_v\) be the width and height of the
axis-aligned bounding box. Define
\(\df{\aes{Aspect Ratio}}=\min\{W,H\}/\max\{W,H\}\).

\subparagraph*{Axis Alignment.}
Grid-like drawings often align vertices on common horizontal or vertical
lines. For \(z\in\{x,y\}\), sort the vertex coordinates
\(z_{(1)}\le\cdots\le z_{(n)}\) and normalize them by
\(\hat z_{(i)}=(z_{(i)}-z_{(1)})/(z_{(n)}-z_{(1)})\).
Scan the normalized values from left to right, starting a new cluster whenever
the next value differs from the first value in the current cluster by more than
\(\varepsilon=10^{-5}\). If the resulting cluster sizes are
\(a_1,\ldots,a_L\), let \(p_i=a_i/n\) and define
\[
S(z)=\frac{n-\bigl(\sum_{i=1}^{L}p_i^2\bigr)^{-1}}{n-1},
\qquad
\df{\aes{Axis Alignment}}=\frac{S(x)+S(y)}{2}.
\]

\subparagraph*{Convexity.}
Convex faces are often easier to read and visually cleaner than non-convex
ones. For each bounded face \(f\), let
\(\mathbf{1}_{\mathrm{conv}}(f)=1\) if all signed turns along its boundary
have the same sign and no three consecutive boundary vertices are collinear,
and let it be \(0\) otherwise. Define
\[
\df{\aes{Convexity}}=
\begin{cases}
1, & |\mathcal{F}|=0,\\
\frac{1}{|\mathcal{F}|}\sum_{f\in\mathcal{F}}
\mathbf{1}_{\mathrm{conv}}(f), & |\mathcal{F}|\ge 1.
\end{cases}
\]

\subparagraph*{Edge-Length Deviation.}
Many drawing styles prefer edges of comparable length. Let
\(\bar{\ell}=m^{-1}\sum_{e\in E}\ell_e\) be the mean edge length and define
\[
\df{\aes{Edge-Length Deviation}}=
\left(1+\frac{1}{m}\sum_{e\in E}
\frac{|\ell_e-\bar{\ell}|}{\bar{\ell}}\right)^{-1}.
\]

\subparagraph*{Edge-Length Ratio.}
The ratio between the shortest and longest edge is an easily interpreted
worst-case measure of edge-length balance. Define
\(\df{\aes{Edge-Length Ratio}}=
\min_{e\in E}\ell_e/\max_{e\in E}\ell_e\).

\subparagraph*{Edge Orthogonality.}
Some drawing styles favor horizontal and vertical edges. For each edge
\(e=\{u,v\}\), let
\(\theta_e=\operatorname{atan2}(y_v-y_u,x_v-x_u)\) and let
\(d_e=\min_{k\in\mathbb{Z}}|\theta_e-k\pi/2|\in[0,\pi/4]\) be its angular
distance from the nearest axis-parallel direction. Define
\[
\df{\aes{Edge Orthogonality}}
=1-\frac{1}{m}\sum_{e\in E}\frac{d_e}{\pi/4}.
\]

\subparagraph*{Face-Area Uniformity.}
A drawing appears more balanced when its bounded faces receive comparable
area. For each bounded face \(f\), let \(A_f\) be the unsigned polygonal area
of its boundary walk and let \(w_f=|f|-2\), the number of triangles in a
triangulation of \(f\). For \(|\mathcal{F}|\le1\), the score is~\(1\).
For \(|\mathcal{F}|\ge2\), define
\(x_f=A_f/\sum_g A_g\) and \(p_f=w_f/\sum_g w_g\), and let
\[
D=\sqrt{\sum_f(x_f-p_f)^2},
\qquad
D_{\max}=\sqrt{1-2\min_f p_f+\sum_f p_f^2}.
\]
With \(D_{\max}\) the largest possible discrepancy, define
\(\df{\aes{Face-Area Uniformity}}=1-D/D_{\max}\).

\subparagraph*{Node Uniformity.}
Vertices should be distributed evenly across the available drawing area.
Partition the bounding box into an \(r\times c\) grid, where
\(r=c=\max\{1,\lfloor\sqrt n\rfloor\}\) and \(T=rc\). If \(n_i\) is the
number of vertices in cell \(i\) and \(\mu=n/T\) is the ideal occupancy, let
\(D=\sum_{i=1}^{T}|n_i-\mu|\). Since the maximum deviation is
\(D_{\max}=2n(T-1)/T\), define
\(\df{\aes{Node Uniformity}}=1-D/D_{\max}\).

\subparagraph*{Spacing Uniformity.}
A drawing should avoid local crowding and highly uneven vertex separation.
For \(n\ge10\), discard the \(\lfloor0.1n\rfloor\) vertices closest to the
boundary of the axis-aligned bounding box; for \(n<10\), retain all vertices.
Let \(U\) be the retained set and, for each \(v\in U\), let
\(\delta_v=\min_{u\in U,\;u\ne v}\|p(u)-p(v)\|_2\). If
\(\bar{\delta}\) and \(\sigma_\delta\) are the mean and population standard
deviation of these nearest-neighbor distances, define
\(\df{\aes{Spacing Uniformity}}=1/(1+\sigma_\delta/\bar{\delta})\).

\subsection{Algorithms}

For the evaluation, we selected algorithms from the graph drawing literature
and practical visualization software. Existing software packages mostly
provide de~Fraysseix--Pach--Pollack or Schnyder-style planar straight-line
layouts, or layouts outside our model, such as orthogonal and polyline drawings
with bends or general-purpose force-directed layouts without planarity
guarantees. We also compare with human-created drawings for a limited subset
of instances from prior studies~\cite{MHWMP25,PAKNPW20}.

Each algorithm takes an abstract planar graph and produces a planar
straight-line drawing. Several algorithms impose additional input
requirements, such as a fixed plane embedding with an outer face or internal 3-connectivity. To satisfy these requirements consistently, we use a common preprocessing pipeline that augments the graph and computes a
combinatorial embedding.
When an algorithm operates on an augmented graph, we restrict its final
drawing to the original graph before computing aesthetic scores.
Since the embedding, outer face, and augmentation can affect the result, using a common scheme improves the comparability and reproducibility of the evaluation; see \cref{sect:eng} for details.

\begin{table}[!tb]
\centering
\caption{Algorithms for producing planar straight-line drawings evaluated in this paper.}
\label{tab:algorithms}
\small
\vfuzz=25pt
\begin{tabular*}{\linewidth}{@{\extracolsep{\fill}}c p{0.18\linewidth}>{\raggedright\arraybackslash}p{0.60\linewidth}p{0.10\linewidth}@{}}
\toprule
& Algorithm & Main idea & Reference \\
\midrule
\multirow{2}{*}{\rotatebox[origin=c]{90}{\scalebox{0.9}{\itshape Grid~}}}
& \alg{Schnyder} & realizer-based grid drawing of a triangulation & \cite{Sch90} \\

& \alg{FPP} & canonical-order shift drawing of a triangulation & \cite{FPP90} \\
\midrule

\multirow{4}{*}{\rotatebox[origin=c]{90}{\shortstack{\itshape Barycentric~~~}}}
& \alg{Tutte} & embedding with fixed convex outer-face boundary & \cite{Tut63} \\

& \alg{CEG-bfs} & boundary-depth weighted embedding for improved spread & \cite{CEG23} \\

& \alg{CEG-xy} & axis-aware weights for horizontal and vertical spread & \cite{CEG23} \\

& \alg{Reweight} & iterative reweighting with weight updates based on face-area imbalance & \cite{FS21} \\
\midrule

\multirow{3}{*}{\rotatebox[origin=c]{90}{\itshape Force-based~~~~}}
& \alg{ForceDir} & force-directed algorithm using node-edge repulsion and rejecting moves that introduce crossings & \cite{FR91,Ber00,GBK24} \\

& \alg{ImPrEd} & force-directed algorithm combining multiple node and edge forces with movement limits and rollback & \cite{SAAB11} \\

& \alg{Air} & pressure-based algorithm that moves vertices using forces derived from neighboring face areas & \cite{Fel14,Kle18b,KMST25} \\
\midrule

\multirow{3}{*}{\rotatebox[origin=c]{90}{\itshape Balancers~~~~}}
& \alg{AngleBalancer} & iterative barycentric-weight optimization to balance vertex angles and penalize very small angles & This work \\

& \alg{EdgeBalancer} & iterative barycentric-weight optimization to balance edge lengths & This work \\

& \alg{FaceBalancer} & iterative barycentric-weight optimization to balance face areas & This work \\
\midrule

& \alg{Hybrid} & multi-method layout selection with postprocessing & This work \\

\bottomrule
\end{tabular*}
\end{table}

\cref{tab:algorithms} summarizes four groups of algorithms; see \cite{PV26} for
the implementation.
The first contains
two combinatorial grid-drawing algorithms. \alg{Schnyder} computes a realizer
and derives combinatorial coordinates, producing a grid-style drawing after
normalization~\cite{Sch90}. \alg{FPP} applies the
de~Fraysseix--Pach--Pollack shift method based on a canonical
ordering~\cite{FPP90}.

The second group uses the classical barycentric approach to produce convex
drawings of internally 3-connected planar graphs. \alg{Tutte} derives
interior-vertex coordinates from fixed outer-face positions by solving a
linear system. \alg{CEG-bfs} and \alg{CEG-xy} of Chiu,
Eppstein, and Goodrich~\cite{CEG23} manipulate barycentric weights for
more spread-out layouts. Inspired by an iterative embedding of Felsner
and Scheucher~\cite{FS21}, \alg{Reweight} repeatedly recomputes the weights
from the current drawing. Edges incident to faces with excessive area receive
larger weights, encouraging them to shorten in the next embedding, thereby
reducing the~face~areas.

The third group consists of local improvement heuristics that add planarity
constraints to classical force-directed layouts~\cite{Kob13,FR91}.
\alg{ForceDir} is motivated by crossing-preserving force-directed
approaches~\cite{Ber00,GBK24}; it rejects moves that introduce crossings and
uses repulsion to encourage uniform vertex spacing. \textsc{ImPrEd} is a more
elaborate implementation of the same high-level idea, combining node-node,
edge, and node-edge forces with sector-based movement limits and explicit
rollback to maintain planarity~\cite{SAAB11}. \textsc{Air} uses pressure-like
local vertex forces derived from incident triangle areas and repeatedly
rebalances them, following the air-pressure viewpoint for prescribed-area
drawings by Felsner and Kleist~\cite{Fel14,Kle18b}.

The last group comprises four new algorithms developed in this work:
three \df{balancer} methods that share an
optimization framework but target different objectives, and \alg{Hybrid},
which uses the aggregate aesthetic score from our evaluation to select and
refine drawings produced by multiple algorithms. We therefore interpret the
results of \alg{Hybrid} separately.

\paragraph*{Balancer Algorithms}

We assume that \(G\) is an internally triangulated plane graph. Following
Tutte's barycentric construction~\cite{Tut63}, we fix its outer face \(B\) as
a convex polygon and place every interior vertex \(u\) at a weighted average
of its neighbors. Specifically, if \(v_1,\ldots,v_{d_u}\) are the neighbors of
\(u\), then \(p_u=\sum_i\lambda_{u,i}p_{v_i}\), where the
\df{barycentric weights} satisfy \(\lambda_{u,i}>0\) and
\(\sum_i\lambda_{u,i}=1\). A larger weight gives the corresponding neighbor
more influence on the position of \(u\).
Solving these equations determines all interior positions and, with positive
weights and a fixed convex boundary, preserves the planar embedding.

The balancers optimize these weights using gradient-based methods.
Directly optimizing the weights would require maintaining their positivity and unit sum. Instead, we introduce an auxiliary unrestricted real parameter
\(\theta_{u,i}\) for each weight and set
\(\lambda_{u,i}=\exp(\theta_{u,i})/\sum_j\exp(\theta_{u,j})\). This
softmax transformation enforces both constraints automatically, allowing an
optimizer to vary the parameters freely. The parameters are initialized from
an initial Tutte drawing. Unlike \alg{Reweight}, which changes weights by a
prescribed local rule,
the balancers adjust all weights jointly to minimize a differentiable objective.

For each parameter choice, we solve the barycentric equations for the vertex
coordinates and evaluate the objective. Since the parameters affect the
objective indirectly through this linear system, we compute the gradient by
solving one additional transposed system; this adjoint method obtains all
parameter derivatives at once instead of solving again for every parameter.
The gradient-based optimizer (L-BFGS) then updates the parameters,
using backtracking to reduce a step until it improves the objective and
preserves the face orientations.
The process stops at convergence or after \(T\) iterations, and the final drawing
is restricted to the original graph. With \(k\) interior vertices, dense
factorization of the \(k\times k\) barycentric system gives \(O(Tk^3)\) time.

The three variants use different objectives. \alg{FaceBalancer} reduces the
variance of bounded triangular-face areas, with penalties for very short or
highly unequal edges. The edge penalties discourage degenerate solutions with
nearly collapsed edges. \alg{AngleBalancer} penalizes deviations of incident
angles at each original vertex from \(2\pi/\deg(v)\), as well as very small
angles. \alg{EdgeBalancer} applies variance, smooth absolute-deviation, and
smooth range terms to the logarithms of the original edge lengths. Using
logarithms makes the \alg{EdgeBalancer} objective invariant under uniform
scaling.
All three methods reject steps that make the triangulated drawing invalid.

\begin{algorithm}[!tb]
    \caption{The \alg{Hybrid} algorithm}
    \label{alg:hybrid-new}
    \small
    \begin{algorithmic}
        \State Let \(\mathcal{A}\) contain the tree, radial-tree, unicyclic, grid,
        outer-cycle, core-tree, \alg{AngleBalancer}, \alg{EdgeBalancer},
        \alg{FaceBalancer}, \alg{Reweight}, \alg{Schnyder}, \alg{CEG-bfs}, and
        \alg{Tutte} methods
        \State Initialize \(\mathcal{C}=\emptyset\)
        \For{each applicable method \(A\in\mathcal{A}\)}
        \State Run \(A\); if it returns a valid drawing \(D\), add \(D\) and its
        best rotation/alignment variant to \(\mathcal{C}\)
        \EndFor
        \State Sort \(\mathcal{C}\) by the mean of the ten metrics, and let
        \(D^\star\) be the best candidate
        \State Let \(k=3\) for \(n\le 50\), \(k=2\) for \(50<n\le 75\),
        \(k=1\) for \(75<n\le 150\), and \(k=0\) otherwise
        \For{each of the top \(k\) candidates \(D\)}
        \State Polish \(D\), reapply rotation/alignment, and update \(D^\star\)
        if \(D\) has a higher score
        \EndFor
        \If{\(n\le 75\)}
        \State Apply finer polishing and convexity repair to \(D^\star\)
        \EndIf
        \State Return \(D^\star\)
    \end{algorithmic}
\end{algorithm}

\paragraph*{Hybrid Algorithm}

For an input graph, we do not know which drawing algorithm
will produce the best layout. Therefore, \alg{Hybrid} applies several of the
evaluated algorithms, together with methods for trees (level and radial
layouts), unicyclic graphs (cycle with attached trees), grids, graphs drawn
from an outer cycle, and cores with attached trees, to generate \df{candidate}
drawings. For each candidate, it searches over rotations and alignment adjustments
that place nearly aligned vertices on common horizontal or vertical lines.
It ranks the resulting drawings by the aggregate aesthetic score (the arithmetic
mean of the ten metrics) and refines the most promising; see
\cref{alg:hybrid-new}.

We call the subsequent local refinement \df{polishing}: each vertex is moved
in eight fixed directions using progressively smaller step sizes, and a move
is retained only if it preserves planarity and improves the aggregate score.
For smaller graphs, \df{convexity repair} additionally moves a reflex
(inward-turning) vertex of a nonconvex face toward the face centroid.

To limit running time on larger graphs, \alg{Hybrid} tries fewer candidate
methods and performs fewer refinement passes. Its size limits, refinement
budgets and step sizes, and alignment tolerances amount to more than thirty
parameters. Their defaults were selected from a few ad hoc configurations by
visually inspecting individual drawings and analyzing aggregate scores; they
were not systematically tuned across the benchmark families.

\section{Evaluation Results}

We first compare the algorithms across the ten aesthetic metrics, then examine
metric tradeoffs and differences among benchmark families. Next, we analyze
implementation failure rates and runtimes, and finally compare automatic and
human-created drawings.

\subsection{Overall Comparison}

\begin{figure}[!tb]
    \centering
    \includegraphics[width=\linewidth]{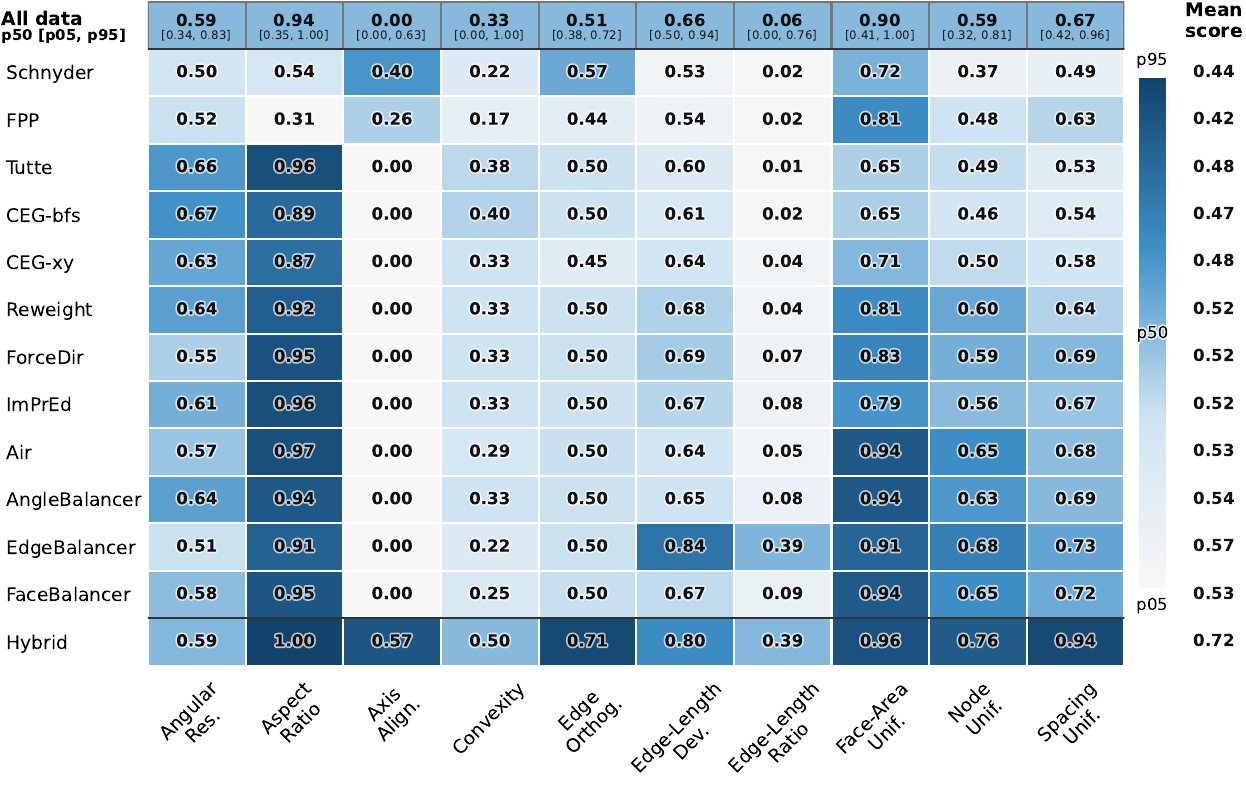}
    \caption{
        For each algorithm and metric, the displayed value is the median over all
        graph instances in the benchmark suite; larger values indicate better
        layouts. The top row reports, for each metric, the median and 5th--95th
        percentile range over all algorithm--instance pairs. Colors are scaled per
        metric, from \(p05\) to \(p95\), with values outside this range clipped.
    }
    \label{fig:metric-rank-heatmap}
\end{figure}

\begin{figure}[!htbp]
    \centering
    \includegraphics[width=\linewidth]{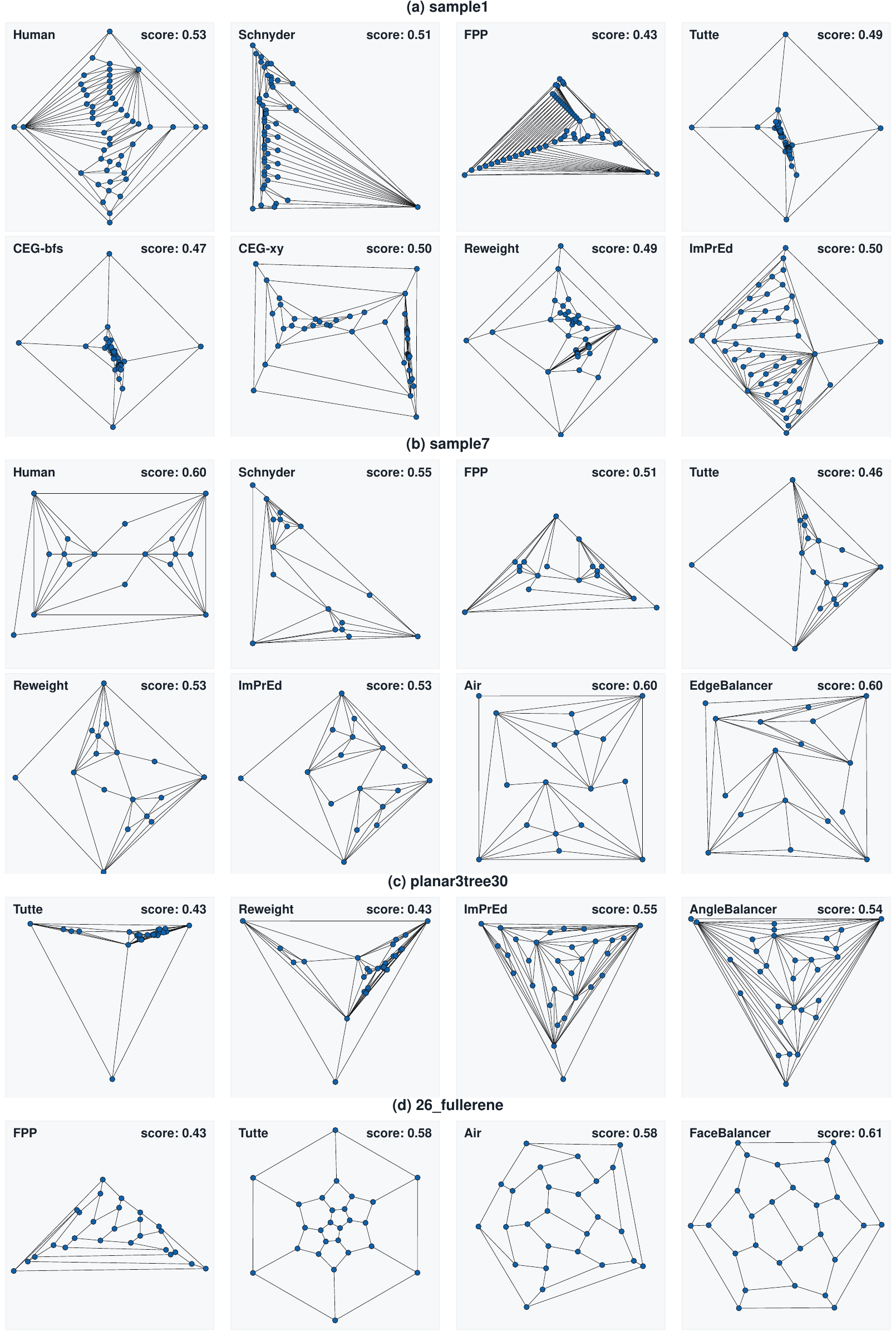}
    \caption{A selection of representative layouts of planar \textsc{Named} graphs produced by different drawing algorithms. \textbf{score} is mean aesthetic value across 10 metrics.}
    \label{fig:samples}
\end{figure}

\cref{fig:metric-rank-heatmap} summarizes the metric values for all algorithms
across the four benchmark collections. Each cell in the table reports, for one
algorithm and metric, the median score over all graph instances; larger scores
indicate better layouts.
The color scale is normalized separately for each metric using its 5th and 95th
percentiles, so that colors emphasize relative differences within a metric rather
than absolute differences across metrics.
The confidence intervals for these median
scores have widths below \(0.001\) in every cell and are therefore omitted.
\cref{fig:samples} shows some resulting drawings. We next analyze
each family of algorithms separately.

Among the individual algorithms, \alg{Schnyder} has the highest median
\aes{Axis Alignment} (\(0.40\)) and \aes{Edge Orthogonality} (\(0.57\))
scores. \alg{FPP} has a median \aes{Face-Area Uniformity} score of \(0.81\),
but the lowest \aes{Aspect Ratio} (\(0.31\)) and \aes{Convexity} (\(0.17\))
scores. Both methods are below the individual-algorithm maxima on
\aes{Angular Resolution}, \aes{Convexity}, and \aes{Edge-Length Ratio}.
\cref{fig:samples}(a,b) illustrates the small angles and non-uniform edge
lengths in the grid-based layouts.

The Tutte-based methods score best on \aes{Convexity} and
\aes{Angular Resolution}. Their \aes{Convexity} scores require caution: the
algorithms draw an augmented graph, whereas the metric evaluates the original
faces, which often become non-convex after dummy vertices are discarded.
Plain \alg{Tutte} has median scores of \(0.96\) on \aes{Aspect Ratio} and
\(0.38\) on \aes{Convexity}, compared with \(0.01\) on
\aes{Edge-Length Ratio}, \(0.65\) on \aes{Face-Area Uniformity}, and
\(0.49\) on \aes{Node Uniformity}, reflecting the uneven scale often produced
by barycentric embeddings. The two \alg{CEG} variants improve different
aspects of this baseline: \alg{CEG-bfs} attains the highest median
\aes{Angular Resolution} score among the individual algorithms (\(0.67\)),
while \alg{CEG-xy} has a higher \aes{Edge-Length Deviation} score than
\alg{CEG-bfs} (\(0.64\) versus \(0.61\)). As expected, \alg{Reweight}
improves \aes{Face-Area Uniformity} over plain \alg{Tutte} but has a median
\aes{Edge-Length Ratio} of \(0.04\), compared with the individual-algorithm
maximum of \(0.39\).
Although easy to implement, these methods remain weak on spacing-oriented
metrics, as illustrated in \cref{fig:samples}(a,c), where large portions of a
graph collapse to a point---a well-known problem for this family of algorithms~\cite{CEG23}.

The three force-directed methods have similar profiles on \aes{Aspect Ratio},
\aes{Axis Alignment}, \aes{Convexity}, \aes{Edge Orthogonality}, and
\aes{Spacing Uniformity}, so choosing among them has limited effect on these
metrics. Within this group, \alg{Air} has the highest median
\aes{Face-Area Uniformity} (\(0.94\)) and \aes{Node Uniformity} (\(0.65\))
scores, \alg{ImPrEd} has the highest \aes{Angular Resolution} score
(\(0.61\)), and \alg{ForceDir} has the highest
\aes{Edge-Length Deviation} score (\(0.69\)).
All three methods have a median \aes{Axis Alignment} score of \(0\), below
the grid methods (\(0.26\) and \(0.40\)), and \aes{Convexity} scores
of \(0.29\)--\(0.33\), below the highest score among the Tutte-based methods
(\(0.40\)).
We emphasize that our implementation of \alg{ImPrEd} is the only runtime outlier among evaluated algorithms; refer to \cref{sect:eng}~for~details.

The balancers improve their target metrics:
\alg{AngleBalancer} has a median \aes{Angular Resolution} score of \(0.64\),
\alg{EdgeBalancer} has the highest scores on both
edge-length metrics (\(0.84\) and \(0.39\)), and \alg{FaceBalancer} ties
for the highest \aes{Face-Area Uniformity} score
(\(0.94\)). These
gains create tradeoffs, as improving one geometric criterion can worsen others,
including \aes{Convexity}, \aes{Axis Alignment}, and
\aes{Angular Resolution}.

Looking across all ten metrics, there is no single winner among the
individual algorithms. The best method depends on the aesthetic criterion:
the top scores are split among methods that optimize different
geometric properties. This observation aligns with the prior
studies~\cite{ADDKL22,MPWK24}. Among the individual algorithms,
\alg{EdgeBalancer} has the best aggregate performance, because it has the
highest individual-algorithm median on both edge-length metrics.

The score-guided \alg{Hybrid} method should be interpreted separately. As
shown in \cref{fig:metric-rank-heatmap}, it obtains the highest aggregate
score and is strong on most metrics, but this is expected because it directly
optimizes the same total score used in the evaluation.
Thus, \alg{Hybrid} serves as a reference point for the metric scores,
rather than a directly comparable winner among drawing algorithms; see
\cref{fig:manual_vs_auto} and  \cite{PV26_gd_collection} for examples of its output.

\subsection{Tradeoffs Between Metrics}

\begin{figure}[!tb]
    \centering
    \captionsetup[subfigure]{justification=centering}
    \begin{subfigure}[b]{.80\linewidth}
        \centering
        \includegraphics[width=\textwidth]{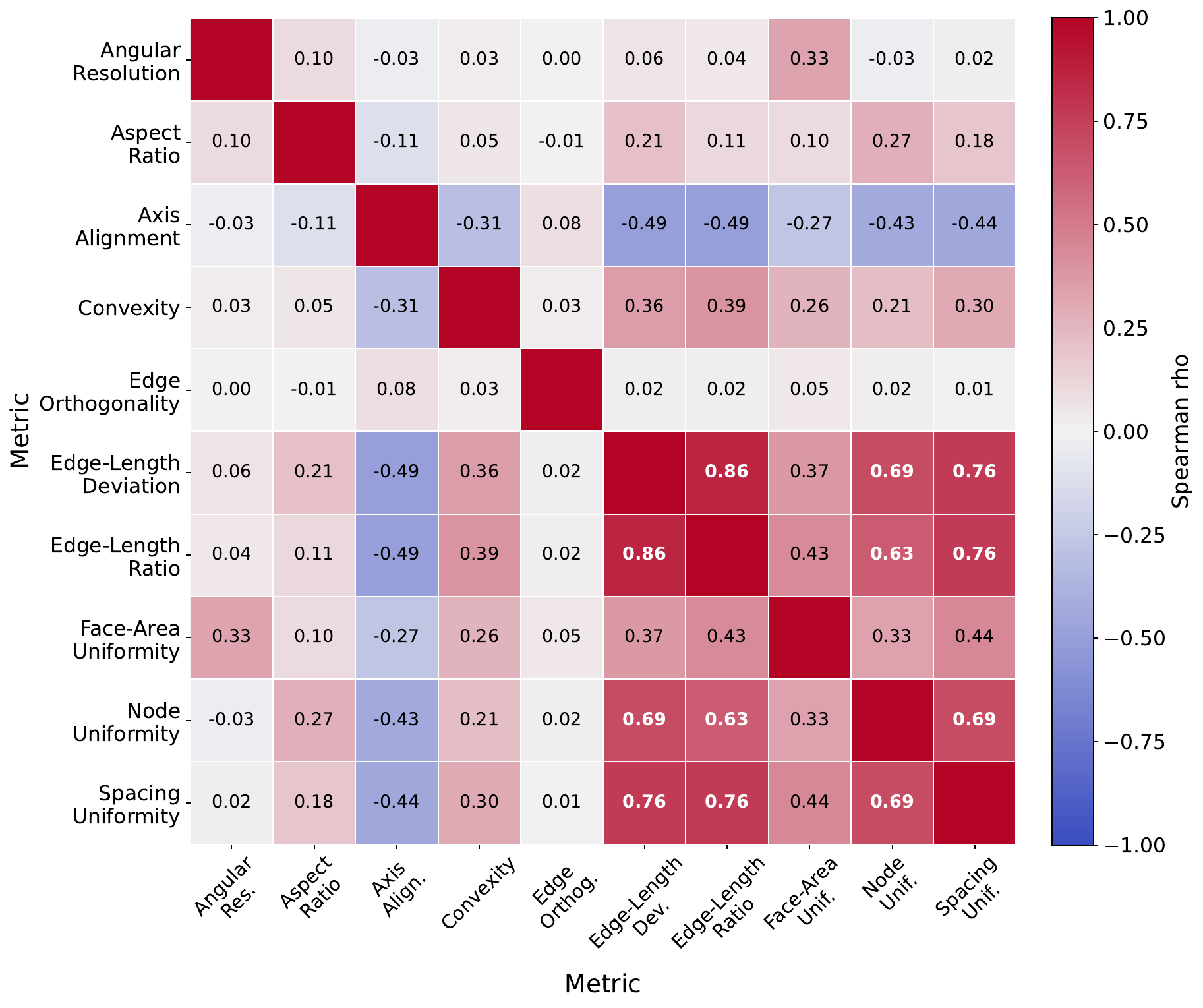}
    \end{subfigure}
    \hfill
    \caption{Spearman correlations between layout metrics over all successful drawings.}
    \label{fig:correlations}
\end{figure}

Prior work argues that graph drawing metrics should be interpreted through their distributions and correlations, and that these relationships depend on benchmark and metric definitions~\cite{MPWK24,MHWMP25}. \cref{fig:correlations} shows the corresponding correlations between aesthetics.

The strongest identified pattern is a cluster of edge-length and
spacing metrics: \aes{Edge-Length Deviation}, \aes{Edge-Length Ratio},
\aes{Spacing Uniformity}, and \aes{Node Uniformity}. In contrast,
\aes{Axis Alignment} is negatively correlated with several of these metrics.
This indicates a recurring tradeoff in our
experiments: layouts with stronger grid-like structure often have less uniform
edge lengths or vertex spacing. These aggregate patterns do
not fully describe individual algorithms. For
example, \alg{Schnyder} has a high median \aes{Axis Alignment} score (\(0.40\))
but a low \aes{Edge-Length Ratio} score (\(0.02\)), while \alg{FPP} has
high \aes{Face-Area Uniformity} (\(0.81\)) and \aes{Spacing Uniformity}
(\(0.63\)), but weak \aes{Aspect Ratio} and \aes{Convexity} scores. Thus, the
global correlation structure summarizes broad metric relationships, but
individual algorithms can realize these tradeoffs differently.

\subsection{Results by Benchmark Family}

We next study whether the aggregate conclusions are stable across benchmark
families. The four families differ substantially in size: \textsc{Rome}
contains 3279 graphs, \textsc{North} 854 graphs, \textsc{GD-collection}
807 graphs, and \textsc{Named} only 48 graphs. Thus, the first three provide
broad benchmark evidence, while \textsc{Named} should be interpreted as a
smaller curated set.

The algorithm rankings by aggregate score are stable across benchmark families, especially
among the three large collections. This stability is visible in
\cref{tab:benchmark-family-rankings}: \alg{Hybrid} and \alg{EdgeBalancer} have
the highest aggregate scores across the benchmarks, while \alg{FPP} and
\alg{Schnyder} remain among the lowest-scoring methods in every family. The
smaller \textsc{Named} set is somewhat less consistent, as expected for a
curated collection of only 48 graphs.

\begin{table}[t]
    \centering
    \caption{Algorithm rankings by aggregate score over all graphs and by
    benchmark family. Scores are arithmetic means over the ten metric values
    within each group.}
    \label{tab:benchmark-family-rankings}
    \small
    \begin{tabular*}{\textwidth}{@{\extracolsep{\fill}}l r l r l r@{}}
        \toprule
        Benchmark & Graphs & \multicolumn{2}{c}{Highest aggregate scores} & \multicolumn{2}{c}{Lowest aggregate scores} \\
        \midrule

        \multirow{3}{*}{\textsc{All}} & \multirow{3}{*}{4988}
        & \alg{Hybrid} & 0.72 & \alg{FPP} & 0.42 \\
        & & \alg{EdgeBalancer} & 0.57 & \alg{Schnyder} & 0.44 \\
        & & \alg{AngleBalancer} & 0.54 & \alg{CEG-bfs} & 0.47 \\
        \addlinespace
        \midrule

        \multirow{3}{*}{\textsc{Rome}} & \multirow{3}{*}{3279}
        & \alg{Hybrid} & 0.71 & \alg{FPP} & 0.41 \\
        & & \alg{EdgeBalancer} & 0.56 & \alg{Schnyder} & 0.43 \\
        & & \alg{AngleBalancer} & 0.53 & \alg{CEG-bfs} & 0.46 \\
        \addlinespace
        \midrule

        \multirow{3}{*}{\textsc{North}} & \multirow{3}{*}{854}
        & \alg{Hybrid} & 0.75 & \alg{FPP} & 0.43 \\
        & & \alg{EdgeBalancer} & 0.58 & \alg{Schnyder} & 0.45 \\
        & & \alg{AngleBalancer} & 0.56 & \alg{CEG-bfs} & 0.50 \\
        \addlinespace
        \midrule

        \multirow{3}{*}{\textsc{GD-collection}} & \multirow{3}{*}{807}
        & \alg{Hybrid} & 0.80 & \alg{FPP} & 0.46 \\
        & & \alg{EdgeBalancer} & 0.68 & \alg{Schnyder} & 0.47 \\
        & & \alg{Reweight} & 0.63 & \alg{CEG-xy} & 0.58 \\
        \addlinespace
        \midrule

        \multirow{3}{*}{\textsc{Named}} & \multirow{3}{*}{48}
        & \alg{Hybrid} & 0.72 & \alg{FPP} & 0.49 \\
        & & \alg{EdgeBalancer} & 0.62 & \alg{Tutte} & 0.54 \\
        & & \alg{FaceBalancer} & 0.61 & \alg{Schnyder} & 0.55 \\

        \bottomrule
    \end{tabular*}
\end{table}

The per-metric picture is similar: \alg{Hybrid} has high observed scores on
most metrics, while metric-specific strengths remain visible, especially
\alg{CEG-bfs} for angular resolution and \alg{EdgeBalancer} for edge-length
metrics. Thus, benchmark choice affects some secondary rankings but not the
main conclusions.

\subsection{Algorithm Engineering}
\label{sect:eng}

Practical planar graph drawing is partly an algorithmic problem and partly an engineering problem. An implementation must not only optimize an aesthetic objective, but also preserve planarity, avoid degenerate geometric cases, and provide reliable defaults for algorithmic parameters.
For example, while the underlying idea for force-directed approaches is simple and well known~\cite{Kob13}, two implementations may produce substantially different outputs;
compare results of \alg{ForceDir} and \alg{ImPrEd} in \cite{PV26_gd_collection}.
Our pipeline uses two common engineering steps.
First, most of the methods run on an augmented plane graph rather than directly on the input graph; this \df{preprocessing} is needed to satisfy algorithm requirements which often impose constraints on the input.
Second, we use a lightweight \df{postprocessing} to improve visual representation
of the result.

\subparagraph*{Preprocessing.}
Several algorithms require stronger input conditions than a general planar
graph, such as a fixed plane embedding, a prescribed outer face, or an internally
triangulated instance.
We therefore apply the same augmentation before running the layout algorithms,
both to satisfy these requirements and to give all applicable algorithms a
common starting point. This is important because the choice of the outer face
can affect both the drawing and its metric values.
Starting from an abstract planar graph, we first compute a plane embedding and
choose the longest face as the outer face. We then stellate each internal face by
adding one dummy vertex adjacent to all vertices of the face, and add a dummy
cycle around the outer face, connecting each original outer-face vertex to its
corresponding dummy vertex. The resulting augmented graph is
internally triangulated and passed to the layout algorithm.
After computing the
layout, we discard all dummy vertices and evaluate the metrics on the original
vertices, edges, and faces.
Algorithms requiring initial coordinates use the plain \alg{Tutte} layout of
the augmented graph as a common initialization. The constructive grid
algorithms, \alg{Schnyder} and \alg{FPP}, do not use this initialization.

We note that the choice of the embedding, the outer face, and the augmentation strategy affect the drawing and the resulting metric
scores. A systematic study of the preprocessing strategy is beyond the scope of
this paper, and we leave it as future work.

\subparagraph*{Postprocessing.}
We also apply lightweight postprocessing to improve
\aes{Axis Alignment} and \aes{Edge Orthogonality} without substantially
changing the overall drawing shape. First, we rotate the drawing to make as
many edges as possible nearly horizontal. As a rigid transformation, this
preserves distances, angles, and planarity. Second, when possible, we move
vertices slightly toward a virtual grid by merging nearby \(x\)- or
\(y\)-coordinates, accepting only moves that preserve planarity. These
operations are deliberately heuristic: they do not replace the layout
algorithm, but make its output more readable.

\begin{table}[!tb]
    \centering
    \caption{Implementation complexity, failure rates, and runtimes over all attempted runs. Implementation complexity is estimated by nonblank, non-comment lines in the code: low \(<300\) lines, medium \(300\)--\(900\) lines, and high \(\ge 900\) lines. Runtime statistics are computed over successful runs only. Here, \(p50\) and \(p95\) denote the 50th (median) and 95th runtime percentiles, respectively.
    The enforced time limit is 300 seconds.}
    \label{tab:robustness-runtime}
    \small
    \begin{tabular*}{\textwidth}{@{\extracolsep{\fill}}l c r r r r@{}}
        \toprule
        \multirow{2}{*}{Algorithm} &
        \multirow{2}{*}{\shortstack{Implementation\\complexity}} &
        \multirow{2}{*}{Failure rate} &
        \multicolumn{3}{c@{}}{Runtime} \\
        \cmidrule(l){4-6}
        & & & \(p50\) & \(p95\) & max \\
        \midrule
        \alg{Schnyder} & \compmedium & 0.00\% & 30 ms & 60 ms & 3.2 s \\
        \alg{FPP} & \compmedium & 0.00\% & 40 ms & 80 ms & 3.5 s \\
        \alg{Tutte} & \complow & 0.22\% & 30 ms & 70 ms & 3.2 s \\
        \alg{CEG-bfs} & \compmedium & 0.46\% & 110 ms & 0.6 s & 274 s \\
        \alg{CEG-xy} & \compmedium & 0.42\% & 120 ms & 0.6 s & 275 s \\
        \alg{Reweight} & \complow & 0.34\% & 150 ms & 0.6 s & 281 s \\
        \alg{ForceDir} & \compmedium & 0.28\% & 0.5 s & 3.1 s & 255 s \\
        \alg{ImPrEd} & \compmedium & 0.18\% & 0.8 s & 17.5 s & 300 s \\
        \alg{Air} & \compmedium & 0.28\% & 0.6 s & 1.6 s & 279 s \\
        \alg{AngleBalancer} & \comphigh & 0.14\% & 0.6 s & 1.7 s & 296 s \\
        \alg{EdgeBalancer} & \comphigh & 0.14\% & 0.2 s & 0.5 s & 75 s \\
        \alg{FaceBalancer} & \comphigh & 0.14\% & 0.2 s & 0.5 s & 82 s \\
        \alg{Hybrid} & \compextreme & 0.00\% & 6.0 s & 20.6 s & 28 s \\
        \bottomrule
    \end{tabular*}
\end{table}

\cref{tab:robustness-runtime} highlights two other practical aspects of
algorithm performance: failure rates and runtimes.
We count a run as failed if it exceeds the time limit or the resulting drawing
contains crossings or degenerate contacts, such as a vertex lying on a
nonincident edge.
Under this definition, the classical
constructive methods are the most robust: \alg{Schnyder} and \alg{FPP} have no
failures, while the other individual algorithms fail on
\(0.1\%\)--\(0.5\%\) of the instances. The zero failure rate of \alg{Hybrid}
has a different meaning: as a score-guided method, it can discard invalid
candidates and fall back to another successful drawing.

Failure modes differ across algorithm families.
For Tutte-based methods, the failures arise from numerical degeneracy.
The barycentric solution can place distinct vertices or edges extremely close
together, causing the validity check to detect an edge contact or a
vertex lying on a nonincident edge.
These effects
are rare but occur in practice. Runs also fail when they exceed the
\(300\)-second limit. For the graph sizes considered in this study,
all individual methods have median runtimes below one second, but
\alg{ImPrEd} has a much heavier tail. The large maximum values in
\cref{tab:robustness-runtime} therefore reflect tail behavior on hard instances
rather than typical performance. For \alg{Hybrid}, expensive candidate methods
and polishing steps are enabled only below size-dependent thresholds, which
keeps its observed maximum runtime low. Since our target use case is graphs
with roughly \(50\)--\(100\) vertices, runtime was not the primary limiting
factor; we focused more on aesthetic results than on low-level optimization.

We also consider implementation complexity as part of the algorithm
evaluation. Most methods fit within a few hundred lines of self-contained
\texttt{JavaScript}; even the balancers use only about one thousand lines. The
harder part is choosing reliable defaults: the augmentation strategy, boundary
conditions, numerical tolerances, step sizes, and objective weights can all
change the output. \alg{Hybrid}, for example, has more than thirty parameters
that affect its behavior, so systematically tuning them across all benchmark
families would require substantial implementation effort.

\subsection{Manual versus Automatic Drawings}
\label{sect:human}

\begin{figure}[!tb]
    \centering
    \includegraphics[width=\linewidth]{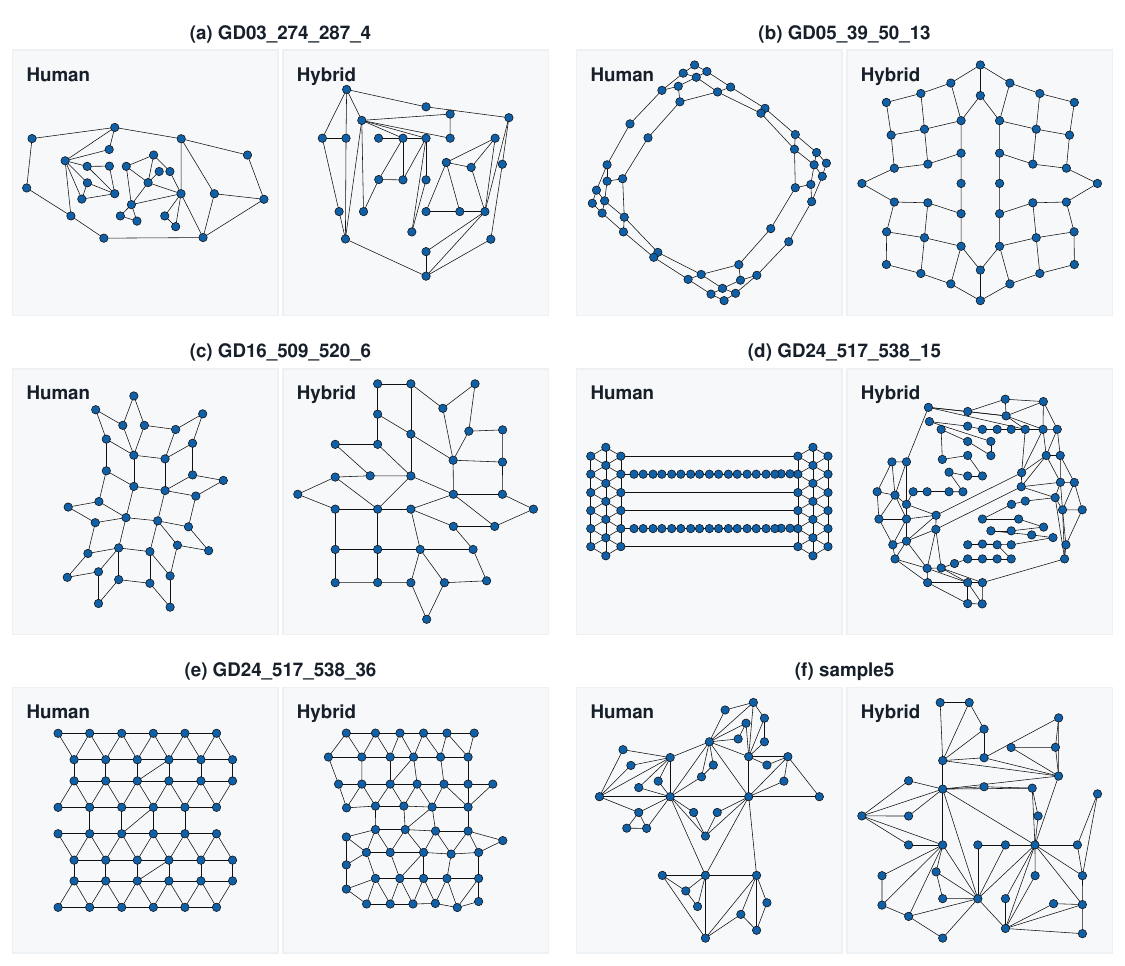}
    \caption{\alg{Human} and \alg{Hybrid} layouts on a selection of instances from
        \textsc{GD-collection} (GD\_*) and \textsc{Named} (sample5) datasets.}
    \label{fig:manual_vs_auto}
\end{figure}

Two benchmark sources provide reference drawings alongside the abstract
graphs. The \textsc{Named} drawings are the human-created layouts used
in~\cite{PAKNPW20}, while the \textsc{GD-collection} drawings are extracted
from published graph drawing papers and may include manual editing or
author-designed layouts. Together, the
sources provide \(646\) reference drawings. We refer to these as \df{human}
drawings to distinguish them from layouts generated by our implementations,
while noting that this designation does not imply that every drawing was produced entirely by hand.
These drawings are evaluated using the same metrics. For each graph and metric, we record the human score
and the highest score achieved by any successful automatic layout. This maximum
is computed independently for every graph and metric, so different algorithms
may provide the best scores for different metrics of the same graph.
\cref{tab:input-vs-best} reports the arithmetic means of the human scores and
these per-graph maxima over all graphs.
Thus, \emph{Best score} is an oracle-style baseline, not the performance of one
fixed algorithm.

\begin{table}[t]
    \centering
    \caption{Comparison of human drawings with the best automatic drawing for each graph and
       metric. The first two numeric columns report mean human and best automatic
        scores.
    The last reports the percentage of graphs for which the human drawing
    scores best among the human and all successful automatic drawings; ties count
    as best.
    }
    \label{tab:input-vs-best}
    \small
    \begin{tabular*}{\textwidth}{@{\extracolsep{\fill}}lrrr@{}}
        \toprule
        Metric & Human score & Best score & Human attains best (\%) \\
        \midrule
        \aes{Angular Resolution}     & 0.58 & 0.69 & 24\% \\
        \aes{Aspect Ratio}           & 0.71 & 0.99 & 8\% \\
        \aes{Axis Alignment}         & 0.47 & 0.62 & 38\% \\
        \aes{Convexity}              & 0.78 & 0.90 & 69\% \\
        \aes{Edge-Length Deviation}  & 0.76 & 0.91 & 9\% \\
        \aes{Edge-Length Ratio}      & 0.34 & 0.68 & 9\% \\
        \aes{Edge Orthogonality}     & 0.61 & 0.73 & 23\% \\
        \aes{Face-Area Uniformity}   & 0.90 & 0.97 & 38\% \\
        \aes{Node Uniformity}        & 0.62 & 0.80 & 12\% \\
        \aes{Spacing Uniformity}     & 0.81 & 0.96 & 12\% \\
        \bottomrule
    \end{tabular*}
\end{table}

\cref{tab:input-vs-best} shows that human drawings are most competitive on
\aes{Convexity}, attaining the best score on \(69\%\) of the graphs, followed
by \aes{Face-Area Uniformity} and the alignment metrics. The largest automatic
gains occur on \aes{Aspect Ratio} and the edge-length metrics,
indicating that the drawings in these benchmark collections have lower scores on balanced aspect ratio and worst-case edge-length balance.

For comparison with a single fixed method, we also compute each drawing's total
score, the arithmetic mean of the ten metrics. Averaged over the \(646\)
graphs, this score is \(0.66\) for \alg{Human} and \(0.78\) for the
score-guided \alg{Hybrid}. This gap may reflect limitations of the aggregate
score rather than show that automatic layouts are preferable.
\cref{fig:manual_vs_auto} illustrates this distinction: in (a), the
\alg{Hybrid} drawing has uneven spacing and several narrow faces, while in
(d), it obscures the long ladder-like structure visible in the human drawing.
This agrees with prior work~\cite{DLFQIRN09}: strong automatic
layouts can match user-generated ones, while manual layouts often make
task-relevant structures more explicit.

\section{Conclusions and Future Work}

Our evaluation suggests two main conclusions. First, no individual algorithm performs best across all aesthetic criteria, and the new balancer methods improve their target
metrics at the cost of other visual properties. Second, \alg{Hybrid} obtains the
strongest aggregate results by combining layout strategies and optimizing a
particular score,
but it still produces visibly suboptimal drawings; see \cref{fig:manual_vs_auto}(a,d) and visit \cite{PV26_gd_collection}. Together with its high implementation complexity, this leaves the following problem open.

\begin{problem}[Automatic planar graph drawing]
    \label{prob:1}
    Given a planar graph, compute a planar straight-line drawing that performs
    well across diverse graph families and aesthetic criteria.
\end{problem}

Such a method should avoid weak scores on individual aesthetics rather than merely maximize an aggregate score. A natural direction is to combine the robustness of constructive algorithms with the flexibility of optimization-based methods. This also requires more principled choices of embedding and outer face, especially for graphs that are not 3-connected.

Progress on \cref{prob:1} requires clearer criteria for evaluating drawings.
The ten aesthetics follow prior work and are useful diagnostics, but do not
fully define drawing quality. They omit properties known to affect perceived
quality, such as symmetry~\cite{DKP18,LHLC18,WK17}. Even for a fixed geometric
property, several scores may be reasonable. Edge-length quality, for example,
may use the worst edge-length ratio, an average or median deviation, or a tail
percentile that ignores outliers. These choices emphasize different defects:
worst-case scores penalize one feature; averages and percentiles describe
global behavior.
A similar issue arises when several aesthetics are combined into one summary score~\cite{HEHL13}. This leads to the second problem.

\begin{problem}[Quality measure for planar drawings]
    Develop a geometric quality measure for planar straight-line drawings that is
    consistent with human judgments.
\end{problem}

Here ``geometric'' means that the measure is based on measurable drawing
properties rather than only a black-box predictor. Human studies report that
perceived quality depends on the algorithm and graph size and generally favors
hand-drawn layouts~\cite{PAKNPW20}. Data-driven pairwise models can outperform
stress and linear combinations of standard metrics, but existing training
labels were based on assumed layout degradation rather than human
judgments~\cite{MPK19}. A natural next step is therefore to collect
human-labeled comparisons and identify the geometric properties that explain
and predict those preferences.

\bibliographystyle{plainurl}
\bibliography{main_planarvibe_arxiv}

\end{document}